\documentclass[12pt]{article} 
\usepackage{psfig}
\usepackage{setspace} 
 \usepackage{bm,amssymb,xspace}

 \usepackage{enumerate} \usepackage{graphicx} \usepackage{latexsym,
 amsmath, theorem, textcomp, dcolumn}

{\theorembodyfont{\rmfamily} }

{\theorembodyfont{\rmfamily} }

\newtheorem{theorem}{Theorem}[section]

\newtheorem{algorithm}[theorem]{\bf Algorithm}

\newcommand{\sR}{\hbox{I\kern-.1667em\hbox{R}}}
\newcommand{\sQ}{\hbox{I\kern-.500em\hbox{Q}}}
\newcommand{\sN}{\hbox{I\kern-.1667em\hbox{N}}}

\renewcommand{\bibitem}{\item[]} \newenvironment{biblist}
   {\begin{list}{} 
    {\setlength{\leftmargin}{4ex}   
     \setlength{\itemindent}{-4ex}  
     \setlength{\itemsep}{0ex}      
     \setlength{\parsep}{0ex}       
   }}{\end{list}}

\renewcommand{\baselinestretch}{1.8}

\begin{document}

\pagestyle{plain}

\title{{\bf Hierarchical Additive Modeling of Nonlinear Association with Spatial Correlations\\-An Application to Relate Alcohol Outlet Density and Neighborhood Assault Rates}\thanks{Supported in part by the National Institute of Alcohol Abuse and Alcoholism Grant NO. R01 AA013810\_03}}


\author{Qingzhao Yu \thanks{Corresponding Author.  Assistant Professor, School of Public Health, Louisiana State University Health Sciences Center.  Email: qyu@lsuhsc.edu}\\
Bin Li  \thanks{Assistant Professor, Department of Experimental Statistics, Louisiana State University.}\\
Richard Scribner \thanks{Professor, School of Public Health, Louisiana State University Health Sciences Center.}\\
Deborah Cohen \thanks{RAND Corporation, Santa Monica, CA.}\\
}

\maketitle \thispagestyle{empty}
\clearpage

\begin{abstract}
\noindent
Previous studies have suggested a link between alcohol outlets and assaultive violence.  In this paper, we explore the effects of alcohol availability on assault crimes at the census tract level over time.  The statistical analysis is challenged by several features of the data: (1) the effects of possible covariates (for example, the alcohol outlet density of each census tract) on the assaultive crime rates may be complex; (2) the covariates may be highly correlated with each other;  (3) there are a lot of missing inputs in the data; and (4) spatial correlations exist in the outcome assaultive crime rates.  We propose a hierarchical additive model, where the nonlinear correlations and the complex interaction effects are modeled using the multiple additive regression trees (MART) and the spatial variances in the assaultive rates that cannot be explained by the specified covariates are smoothed trough the Conditional Autoregressive (CAR) model. We develop a two-stage algorithm that connect the non-parametric trees with CAR to look for important variables covariates associated with the assaultive crime rates, while taking account of the spatial correlations among adjacent census tracts.  The proposed methods are applied to the Los Angeles assaultive data (1990-1999) and compared with traditional method.
\vspace{.25in}

\end{abstract}

\noindent
{\bf Keywords:} Alcohol related crimes; Backfitting; Conditionally autoregressive (CAR) model; Multiple additive regression trees (MART), Nonparametric regression.



\noindent

\noindent

\pagebreak

\section{Introduction}\label{intro}

The alcohol related crime research is important in that it helps governments balance the competing interests of the alcohol industry to increase the distribution and consumption of alcohol and public safety to minimize risks associated with increasing the physical and social availability of alcohol.  There are a large number of published empirical observations of direct relationships between alcohol outlets and measures of interpersonal violence.  Since the effect of alcohol outlets on violence is believed to be contextual, these analysis involve areal data that require more sophisticated techniques to account for their spatial and temporal structure.  Initial studies on the role of alcohol outlets in the neighborhood environment and assaultive violence were ecological in design and conducted at the city level (Scribner et al. 1995; MacKinnon et al. 1995, Watts and Rabow 1983).  As geographical information systems (GIS) software became widely available, more local units of analysis (e.g. census tracts) were used when modeling the theoretical relation between alcohol outlets in a neighborhood and assaultive violence (Scribner et al. 1999; Speer et al. 1998).  Accounting for the spatial autocorrelation, i.e. the possible spillover relation between outlets and assaultive violence into contiguous neighborhoods, was a subsequent advancement (Gorman et al. 2001; Gruenewald 2000; Zhu et al. 2004; Yu et al. 2007).  Recently, longitudinal models are being incorporated into the analysis of the data, introducing an additional level of complexity (Gruenewald and Remer 2006).  In this study we apply a hierarchical additive model to explore possible coefficients that are related to changes in assaultive violence rates among census tracts affected by the 1992 Civil Unrest in Los Angeles which resulted in the immediate loss of over 250 alcohol outlets, and the permanent loss of roughly 150.  

Several features of the analysis present substantial statistical challenges.  First, the response variable, i.e. the assault rates, may be nonlinearly dependent on the covariates. There are two common approaches to deal with nonlinearity: (1) transformation and (2) basis expansion (such as using spline basis). However, the former are not readily applicable when there are a large number of covariates, while the latter requires basis specification (e.g. specify the number and locations of knots).  Moreover, if we apply complicated nonparametric models (some black boxes), we meet the challenge of model interpretation.  For example, it is difficult to interpret the importance and marginal effect for each covariate in the model.   Second, complicated interactions might exist among covariates. Third, more than 7\% of the observations have missing inputs. Finally, in the analysis, we should take into account spatial correlations with adjacent tracts.  A handy method to deal with the correlation is to use a hierarchical model where the spatial correlations are modeled through correlated spatial errors.  We are consequently challenged by combining the hierarchical structure of spatial errors with nonlinear association modeling. 
      
Multiple additive regression trees (MART) is a tree-based ensemble method developed by Friedman (2001). Empirical results have shown that MART achieves highly accurate prediction performance comparing to its competitors. Moreover, comparing to the classical parametric regression methods, MART has the following advantages: (1) MART is able to capitalize on the nonlinear relationships between the dependent and independent variables with no need of specifying the basic functions. Unlike many automated learning procedures, which lack interpretability and operate as a ``black box'', MART provides great interpretation tools (see, for example, relative variable importance and partial dependence plot in Section~\ref{sec-interpretation}). (2) Due to the hierarchical splitting scheme in regression trees, MART is able to capture complex and/or high order interaction effects. (3) As a tree-based method, MART can handle mixed-type predictors (i.e. quantitative and qualitative covariates) and missing values in covariates.  Hence, to handle the first three challenges in analyzing our data, we could use MART.

To tackle the last challenge, we propose a two-stage iterative algorithm to build hierarchical additive models.  At the first stage, a MART model is built to explore the associations between the smoothed assault rates and the covariates.  Here the {\it smoothed assault rates} refer to the original assault rates minus the estimated spatial errors obtained from the second stage.  At the second stage, the spatial correlations in assault rate that could not be explained by covariates are ``smoothed'' through the conditional autoregressive model (CAR).  The two stages iterate until convergence, whose condition is described in Section~\ref{method}.  Our algorithm is an extension of backfitting process (Hastie and Tibshirani, 2003) to more complicated nonparametric settings.  In this article, we apply the hierarchical additive modeling strategy to evaluate the association between the alcohol availability and assault rates in some census tracts in Los Angelos from the year 1990 to 1999.

The rest of the article is organized as follows.  We describe in Section~\ref{data} the data and environment.  We present the two-stage hierarchical additive model in Section~\ref{method}.  In Section~\ref{results}, we apply the hierarchical additive model to analyze the data and compare the results with those from hierarchical linear regression model.  Concluding remarks and future researches are given in Section~\ref{conclusion}.

\section{The 1992 Civil Unrest and Data}\label{data}

\subsection{The 1992 Civil Unrest}	

Our study is designed to capitalize on a natural experiment.  The experiment was made possible by the civil unrest in Los Angeles that followed the verdicts acquitting the police officers accused of beating Rodney King.  The riots resulted in 53 deaths, 2,325 reported injuries, more than 600 buildings completely destroyed by fire, and approximately \$735 million in total damages (Evans 1993).  Many of the buildings destroyed were serving as alcohol purchase outlets.  As a result, a total of 279 liquor licenses were surrendered across 144 census tracts in Los Angeles County due to interruption of their services.  In the wake of the civil unrest, an effort to halt the rebuilding of off-sale alcohol outlets was successful in restricting the re-licensing of outlets with a history of problems (e.g., assaults, homicides, drug sales) around their premises.  The effort was responsible, in part, for over 150 outlets permanently closing in the civil unrest area.  These events provide a natural experiment setting to test various hypotheses regarding the effect of closure of off-sale liquor outlets in 144 tracts, compared to 336 tracts also exposed to the civil unrest where outlets were not affected (Cohen et al. 2006). 

The 1992 civil unrest occurred over a large area of South Central Los Angeles.  In the present study we include only those census tracts in the area affected by the civil unrest, thereby controlling for a possible global effect of the unrest on outcomes.  That is, any measured effect associated with specific temporal changes in a tract's neighborhood environment should be independent of any global effect of the civil unrest, which would affect all tracts regardless of the presence or absence of changes in their neighborhood alcohol environment.  To define the study area in this manner, we used the definition established by the Rose Institute of California State and Local Government at Claremont McKenna College (http://ccdl.libraries.claremont.edu/col/ric/) to study the economic impact of the civil unrest (Hubler 2002).  A total of 480 census tracts comprise the unrest area.  These tracts contained 2,641,320 people in 1990, of whom 48\% were Hispanic and 27\% were African American.  A total of 2,240 unique addresses were damaged in the 480 tracts, while 144 tracts had one or more off-sale liquor outlets whose license was surrendered.  The majority of the damaged addresses were commercial businesses.  Immediately following the civil unrest, there emerged a grass roots effort among the affected communities to halt the rebuilding of alcohol outlets based on the finding that an over-concentration of off-sale outlets existed in the low socioeconomic status areas prior to the civil unrest (Grills et al., 1996; Berestein, 1994; Kang, 1994).   

\subsection{Data}  

The study time frame is 1990 to 1999.  

\textbf{Assaultive Violence}\hspace{1cm}
Our measure of assaultive violence was obtained from the Los Angeles Police Department.  Uniform Crime Report (UCR) offenses involving assaultive violence (i.e., murder, rape, robbery, and assault) were obtained for the years 1990 through 1999.  A summary measure of the count of all violent offenses was generated for each census tract for all study years by geocoding the data that contained the street address of the offense location.   

\textbf{Alcohol Outlet Density}\hspace{1cm}
Measures of alcohol exposure included surrender of off-sale liquor outlet license following the May, 1992 civil unrest, the percentage of off-sale liquor outlet licenses surrendered, and annual off-sale outlet density from 1990 to 1999.  Annual counts of liquor outlet licenses came from the California Department of Alcohol Beverage Control (ABC).  A list of outlets that surrendered their licenses and a list of stores with riot damage as a result of the 1992 Los Angeles civil unrest were also obtained from the ABC.  Alcohol outlets were classified based on their license to sell alcohol for on-premise (bars and restaurants) or off-premise (liquor stores, grocery stores, and convenience stores) consumption using license codes provided by the ABC.  All unique address listings were geo-coded and mapped to the 1990 Census tract areas, and individual data sources were matched by census tract.  Ninety-eight percent of addresses were matched using Arcview 3.2 GIS software (ESRI Inc, Redlands, CA) along with Los Angeles County Topographically Integrated Geographic Encoding and Referencing (TIGER) street files from the 2000 census.  Addresses that the computer was unable to match were hand placed with the help of an Internet mapping site (Mapquest) and a Thomas Guide map book.  

\textbf{Additional Covariates}\hspace{1cm}
Additional tract-level covariates included in the analysis were (1) percentage African American, (2) percentage Hispanic, (3) percentage male between the ages of 15 and 30 years, (4) extent of physical damage in the census tract, and (5) population density.  The first four covariates are annual estimates available for the years 1990-1999 and are included to control for changes in tract composition over time, an endogenous change that could explain temporal changes in assaultive violence rates.  For example, it is possible that changes in assaultive violence are the result of the movement of populations at higher or lower risk for violence into or out of particular study tracts over the course of the observation period.  

The annual estimates of population distributions by age, race and sex were obtained from the Los Angeles County Department of Health Services, with actual counts available for 1990 and 1995 and counts for the other years estimated from birth and death records.  The remaining socio-demographic data were obtained from the 1990 U.S. Census data of Los Angeles County.  Information on damaged buildings (Ong 1993) came, directly or indirectly, from four different sources: the Los Angeles City Department of Building and Safety, the Korean Central Daily, the California Insurance Commission, and the Compton Department of Building and Safety.  Physical damage is measured as a binary indicator of any damage of property in the tract due to the civil unrest.  We also derived a measure of damage density (i.e. damage per square mile), calculated as the ratio of the number of unique addresses damaged in the 1992 civil unrest to the amount of land in the tract used for commercial purposes.  The denominator corresponds to land used for commercial purposes because most of the damaged property was commercial.  To compute the denominator, we estimated the proportion of commercial space in the tract using a land use file and multiplied it by the area of the tract in 1990 in square miles.

\section{Hierarchical Additive Modeling}\label{method}

In this section, we first review the basic ideas under MART and CAR and then we propose a hierarchical additive model - the related algorithm in model building and how to explain the models.

\subsection{Multiple Additive Regression Trees}\label{sec-mart}
MART is a special case of the generic gradient boosting approach developed by Friedman (2001). Given $n$ observations of the form $\{y_i,\mathbf{x}_i\}_1^n=\{y_i,x_{i1},\ldots,x_{ip}\}_1^n$ and any differentiable loss function $L(y,F(\mathbf{x}))$. MART considers the common problem of finding a function $F(\mathbf{x})$ mapping a $p$ dimensional input vector $\mathbf{x}$ to response variable $y$, such that over the joint distribution of all $(y,\mathbf{x})$ values, the expected value of the loss function $L(y,F(\mathbf{x}))$ is minimized. MART approximates the target function $F(\mathbf{x})$ by an additive expansion of trees 
\begin{eqnarray}
\hat{f}(x)=\sum_{m=1}^M \nu b_H(\mathbf{x};\mathbf{\gamma}_m), \label{eq-mart}
\end{eqnarray}
where $b_H(\mathbf{x};\mathbf{\gamma}_m)$ is an $H$-terminal node tree (which partitions the input space into $H$-disjoint regions); $\mathbf{\gamma}_m$ is the parameter vector in building tree m and $\nu \in (0,1)$ is the `shrinkage' parameter ($0<\nu \le 1$) which controls the \textit{learning rate} of the procedure. Empirical results have shown (see e.g., Friedman, 2001, Friedman and Meulman, 2003) that small values of $\nu$ \textit{always} lead to smaller generalization error. The detailed algorithm of MART (for regression) is the following. 
 
\begin{algorithm}\label{alg-mart} MART Algorithm (Friedman 2001)
\begin{enumerate}
\item $\hat{f}_0(\mathbf{x})=arg\displaystyle\min_{\gamma} \sum_{i=1}^n L(y_i,\gamma)$.
\item Repeat for $m=1,2,\ldots,M$:
\begin{enumerate}
\item $\tilde{y}_i=-\left [ \frac{\partial L(y_i, f(\mathbf{x}_i))}{\partial f(\mathbf{x}_i)}\right ]_{f(\mathbf{x})=\hat{f}_{m-1}(\mathbf{x})}, \, i=1,2,\ldots,n$.
\item $\{R_{hm}\}_1^H=H$-terminal node tree on $\{\tilde{y}_{im},\mathbf{x}_i\}_1^n$. 
\item $\gamma_{hm}=arg \displaystyle\min_{\gamma} \sum_{\mathbf{x}_i \in R_{hm}} L(y_i,\hat{f}_{m-1}(\mathbf{x}_i)+\gamma)$.
\item $\hat{f}_m=\hat{f}_{m-1}+\nu \cdot \gamma_{hm} I(\mathbf{x} \in R_{hm})$. 
\end{enumerate}
\item End algorithm.
\end{enumerate}
\end{algorithm}

Within each iteration $m$, a regression tree, whose splitting scheme $\{R_{hm}\}_{h=1}^H$ is optimized based on the negative gradient $\{\tilde{y}_i\}$ at its current estimate $\hat{f}_{m-1}$ (which is closely related to the \textit{steepest-descent} minimization approach in function optimization), is fitted with an estimate $\gamma_{hm}$ in each region. The value of $M$, i.e. the number of iterations or trees, can be chosen based on either cross-validation or monitoring the prediction performance on ``out-of-bag'' samples with subsampling in each iteration (see Friedman, 2002). Note that in practice, we can pre-specify the maximum depth $D$ for individual tree instead of the number of terminal nodes $H$. For example, the tree with $D=1$ (single-split trees with only two terminal nodes) fits an additive model without interaction and MART-fitted model with $D=3$ is able to account for at most three-way interactions. For details of MART and gradient boosting, we refer the readers to the original paper by Friedman (2001). In this paper, MART is run by using the \textbf{gbm} package in \textbf{R}, produced by Greg Ridgeway. 
           
\subsection{Conditional Autoregressive Model}

We use the vector $\{\phi_{T_i,C_i}\}$ to capture spatial autocorrelations among areas $C_i$ at time $T_i$, where $i=1,\ldots,n$ and $n$ is the total number of observations; $T_i = 1,\ldots, T,$ where $T$ is the total number of time slots and $C_i = 1, \ldots, C$ where $C$ is the total number of locations.  A popular model for the spatial correlation arises by assuming that an area $C_i$ is correlated with only the areas that are adjacent to it.  Let $f(\mathbf{x})$, a function of the covariate vector $\mathbf{x}$, explores the association between $\mathbf{x}$ and the response variable $y$.  We have the following model for y with a hierarchical structure on its mean function:
\begin{eqnarray}
y_i &\sim& N(\mu_i,\sigma^2) \mbox{ and } \mu_i = f(\mathbf{x}_i)+\phi_{T_i,C_i};	\label{CAR}	
\end{eqnarray}
\noindent
where $y_i$ is the observed value of the response variable in the area $C_i$ at time $T_i$.  We apply a conditional autoregressive (CAR; Besag 1974) structure for the spatial term $\phi_{T_i,C_i}$.  Let $j\sim i$ denote adjacency of regions $i$ and $j$, and $n_j$ be the number of tracts adjacent to tract $j$.  The hierarchical CAR structure for $\{\phi_{T_i,C_i}\}$ has the form
\begin{eqnarray}
\phi_{T_i,C_i=j}|\phi_{T_i,C_i\neq j}\sim N\left( \sum_{k\sim j}\frac{1}{n_j}\phi_{T_i,k},\frac{1}{n_j\tau_{T_i}}\right);\label{CAR2}
\end{eqnarray}
\noindent
where $\tau_{T_i}$ is the precision parameter controlling the degree of spatial smoothing in $T_i$.  To employ the Winbugs software to find the MLE of $\phi$, we use Equation~\ref{CAR2} as the prior distribution for $\phi$ and also, we assign the temporal smoothing term $\tau_{T_i}$ and the random precision term $1/\sigma^2$ a noninformative uniform hyper distribution ranging from 0 to $\infty$.  In this case, the MLEs for the spatial terms are obtained at the modes of their posterior distributions in terms of Bayesian analysis.  We obtain posterior distributions for all parameters of interest via Markov Chain Monte Carlo (MCMC) algorithms implemented in WinBUGS (free software available at http://www.mrc-bsu.cam.ac.uk/bugs/welcome.shtml).  Notice that we can also obtain MLE for $\phi$ through other iterative optimization algorithms.  

\subsection{The Two-Stage Iteration Algorithm}

To combine the MART, which explores the variable relationship, and the CAR, which identifies a spatial autoregressive structure, we use a two-stage iteration algorithm.  In the first stage, we explore the important covariates.  In the second stage, we smooth the spatial correlations that cannot be explained by the covariates.  To avoid redundancy, the observations in those time slots that the covariates can explain most of the spatial correlation in the response variables will be imported to the second stage for further analysis.  To test whether spatial correlations remained in $y$, we use the test statistics \textit{Moran's I} (Banerjee et al., 2003).  We assume an additive structure between $f(\mathbf{x})$, the effects on $y$ from the covariate $\mathbf{x}$, and the remained spatial correlations.  In the algorithm, $\mathbf{x}$ is the vector of possible covariates; $q$ counts the number of iterations; $\Delta$ is a small constant set beforehand to control convergence and $\delta$ is used to test convergence, measuring the relative difference in the expected values of the responses from the last iteration to current iteration.

\begin{algorithm}\label{two-stage} Two-Stage Analysis
\begin{enumerate}
\item Let $\phi_{T_i, C_i}^0 = 0$ where $C_i\in\{1,\ldots,C\}$, $T_i\in\{1,\ldots,T\}$; $q=0$, $\delta=1000$, $\mu_{1i}=0$ and $i=1,\ldots,n$.
\item If $\delta<\Delta$, go to step~\ref{end}), otherwise q=q+1 and \label{iter}
 \begin{enumerate}
 \item Let $yz_i=y_i-\phi_{T_i, C_i}^{[q-1]}$.  Fit MART $f^{[q]}(\mathbf{x})$ where the response variable is $yz$ and the covariates are $\mathbf{x}$.
 \item Let $e_i=y_i-f^{[q]}(\mathbf{x}_i)$, calculate the Moran's I of $e_i$ for each time slot $T_i$.  Let $S$ be the collection of time slots in which the spatial correlation test show a p-value smaller than $0.01$. 
 \item If $S$ is empty, let $\phi_{T_i, C_i}^{[q]}=0$ and go to step~\ref{end}); otherwise \\
using only the observations $i\in\{i: T_i\in S\}$, let the $f(\mathbf{x}_i)$ in Equation~(\ref{CAR}) be $f^{[q]}(\mathbf{x}_i)$ and calculate the MLEs of $\hat{\phi}_{T_i, C_i}$.  Let $\phi_{T_i\in S, C_i}^{[q]}=\hat{\phi}_{T_i, C_i}$ and $\phi_{T_i\notin S, C_i}^{[q]}=0$.
 \item Let $\mu_{0i}=\mu_{1i}$, $\mu_{1i}=f^{[q]}(\mathbf{x}_i)+\phi_{T_i, C_i}^{[q]}$ and let $\delta=\frac{\sum_{i=1}^{n}(\mu_{1i}-\mu_{0i})^2}{\sum_{i=1}^{n}\mu_{1i}^2}$, go back to~\ref{iter}). 
\end{enumerate}
\item Output the results from step $q$.\label{end}
\end{enumerate}
\end{algorithm}

  Little is known of the convergence property of the above procedure.  In proposing the ICM (Iterated Conditional Modes), Besag (1986) discussed the complication in parameter estimation when there are other parameters to be estimated besides the variance structure.  Ideally, we estimate $f(\mathbf{x})$ and $\phi$ from training data alone and to use the estimated values in subsequent reconstruction.  But no training data are available and it is necessary to estimate $f(\mathbf{x})$ and $\phi$ as part of the restoration procedure.  Meng and Rubin (1993) showed that under certain conditions, iterative conditional maximization converges to local maximizers.  But the problem is even more complicated here as we need to estimate the whole function (build multiple trees) rather than a few parameters.  Our algorithm is essentially a backfitting process (Hastie and Tibshirani, 2000) with the MART and the CAR variance structure as two additive components.  Buja et al. (1989) proved the convergence of the backfitting process for a certain class of fixed, nonadaptive operators.  And the algorithm seems well behaved in general (Hastie and Tibshirani, 2000).  In our analysis, we let $\Delta=10^{-7}$.  That is, if the relative difference  $\delta$ in response means between two sequent iterations is less than $10^{-7}$, we conclude that the algorithm converges.

\subsection{Interpretation}\label{sec-interpretation}
Among the most important ingredients in any interpretation are identifying which variables are important for prediction, and understanding their joint effect on the response.  For tree-based methods, Breiman {\it et al.} (1984) proposed a measure of importance $I^2_j(b_H)$ for each variable $x_j$, based on the number of times that variable was selected for splitting in the tree $b_H$ weighted by the squared improvement to the model as a result of each of those splits. Friedman (2001) generalized this importance measure to additive tree expansions by taking the average over the trees
\begin{eqnarray}
I^2_j=\frac{1}{M}\sum_{m=1}^MI_j^2(b_H(\mathbf{x},\gamma_m)). \label{eq-pd3}
\end{eqnarray}
The measure~(\ref{eq-pd3}) turns out to be more reliable than a single tree as it is stabilized by averaging.  Since these measures are relative, we scale the measure so that the importance of all the variables sum to 100\%.  

In addition to the importance measure, Friedman (2001) also introduced a concept called {\it partial dependence} to describe the dependence of the fitted model on a subset of variables. Given any subset $\mathbf{x}_s$ of the input variables indexed by $s \subset \{1,\ldots,p\}$. The partial dependence is defined as 
\begin{eqnarray}
F_s(\mathbf{x}_s)=E_{\mathbf{x}_{\setminus s}}[f(\mathbf{x})], \label{eq-pd1}
\end{eqnarray}
where $E_{\mathbf{x}_{\setminus s}}[\cdot]$ means expectation over the joint distribution of all the input variables with index not in $s$.  In practice, partial dependence can be estimated from the data by 
\begin{eqnarray}
\hat{F}_s(\mathbf{x}_s)=\frac{1}{n}\sum_{i=1}^n\hat{f}(\mathbf{x}_s,\mathbf{x}_{i\setminus s}), \label{eq-pd2}
\end{eqnarray}
where $\{\mathbf{x}_{i\setminus s}\}_1^n$ are the data values of $\mathbf{x}_{\setminus s}$. 

To explore the spatial autocorrelations, we could draw a map with the MLEs of spatial residuals.  The spatial residuals are residuals that are spatially correlated and cannot be explained by covariates in the model.  The map of spatial residuals could suggest various spatially-varying covariates that are still missing in the model.

\section{The Hierarchical Additive Model Application To The Alcohol Related Crime Study}\label{results}
 
In this section, we use the hierarchical additive model developed in Section~\ref{method} to analyze the data introduced in Section~\ref{data}.   The purpose of the analysis is to find out whether there is an association between alcohol availability and assaults.  In this analysis, the response variable is defined to be $y_i=\log\left[\frac{\mbox{(number of assault)}_i \mbox{+ 0.0001}}{\mbox{population}_i}\times 1000\right]$, the assault rate per 1000 people. We add $0.0001$ to the number of assault to avoid the complication so that the term in the log scale is always larger than 0.  The area unit in this analysis is census tract, ranging from 1 to 290 and the time unit is year.  

We have 10 years data from 1990 to 1999.  An exploratory analysis suggests that we use the possible covariates: 1) heterogeneity in race - the covariates represent the proportion of residents that are black, white, Asian and Hispanic (the variable names in the model are ``black'', ``white'', asian'' and ``hispanic''); 2) percentage of young males in neighborhood (``male\_15\_30''); 3) percentage of households in poverty (``poverty''); 4) damage level measuring the relative damage caused by civil unrest - the covariate is called ``damage'', calculated as the number of damaged addresses in the tract divided by the tract area in square miles; 5) the years (``year''); and 6) the alcohol availability measurements, including the on-premise, off-premise, total alcohol outlet density and the indicator variables of the alcohol license surrender during the civil unrest.  

\subsection{Relative Variable Importance And Partial Dependence}   

We fitted the following two models.  Model A uses the variable ``totaldensity'', which is defined as the number of on-premise and off-premise alcohol outlets per roadway mile, to control the alcohol availability.  We also include the variables ``pctonsurryn'' and ``pctoffsurryn'', the percentages of on-premise and off-premise alcohol licenses surrendered in the 1992 civil unrest separately, to check if a sudden decrease in alcohol outlets would lead to important assault rate changes.  In model B, we use ``onsale'' and ``offsale'', the on-premise and off-premise alcohol outlet densities per roadway mile, to check whether the on-premise or the off-premise alcohol availability is a more important variable in predicting assault rate.  In this model, we use two indicator variables ``onsurryn'' and ``offsurryn'', to indicate whether there is on-premise or off-premise alcohol license surrender in the corresponding census tract in the 1992 civil unrest.  In MART, we set the learning rate $\nu$ at 0.001 and the maximum depth for each individual tree at 3, i.e. model counts up to three-way interactions.  

\begin{figure}[htb]
 \centering
 \includegraphics[angle=270,width=1\textwidth]{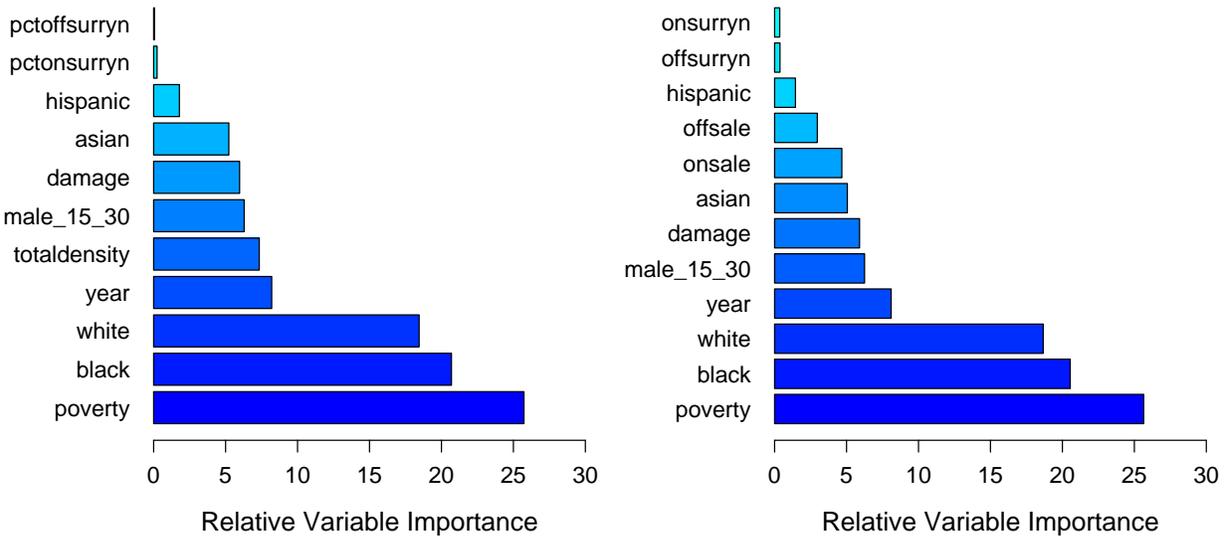}
 \caption{\it{Relative variable importance in MART-fitted model A (left) and B (right).}}\label{fig-imp}
\end{figure}

It is useful to understand the relative importance or contribution of each explanatory variable to the fitted model. Figure~\ref{fig-imp} shows the relative variable importance in MART-fitted models in Model A (left) and B (right), which consist of 4780 and 4519 individual trees, respectively. In Figure~\ref{fig-imp}, we see that compared with other covariates in the model, the variable ``poverty'' is the most important variable in predicting assault rates in both models.  From the left panel of Figure~\ref{fig-imp}, we find that alcohol availability (``totaldensity'') is a relative important variable, ranked at the fifth in the model, more important than the proportion of young males, the proportion of Asians and Hispanics and the damage level in the Civil Unrest.  The percentages of on-premise or off-premise license surrendered in the civil unrest are relatively unimportant in the model.   When we consider the effect of on-premise and off-premise alcohol availabilities separately in model B, we find on-premise alcohol outlet density is a little bit more important than that of the off-premise (the right panel of Figure~\ref{fig-imp}) in explaining the assault rates.  

\begin{figure}[htb]
 \centering
 \includegraphics[angle=270,width=1\textwidth]{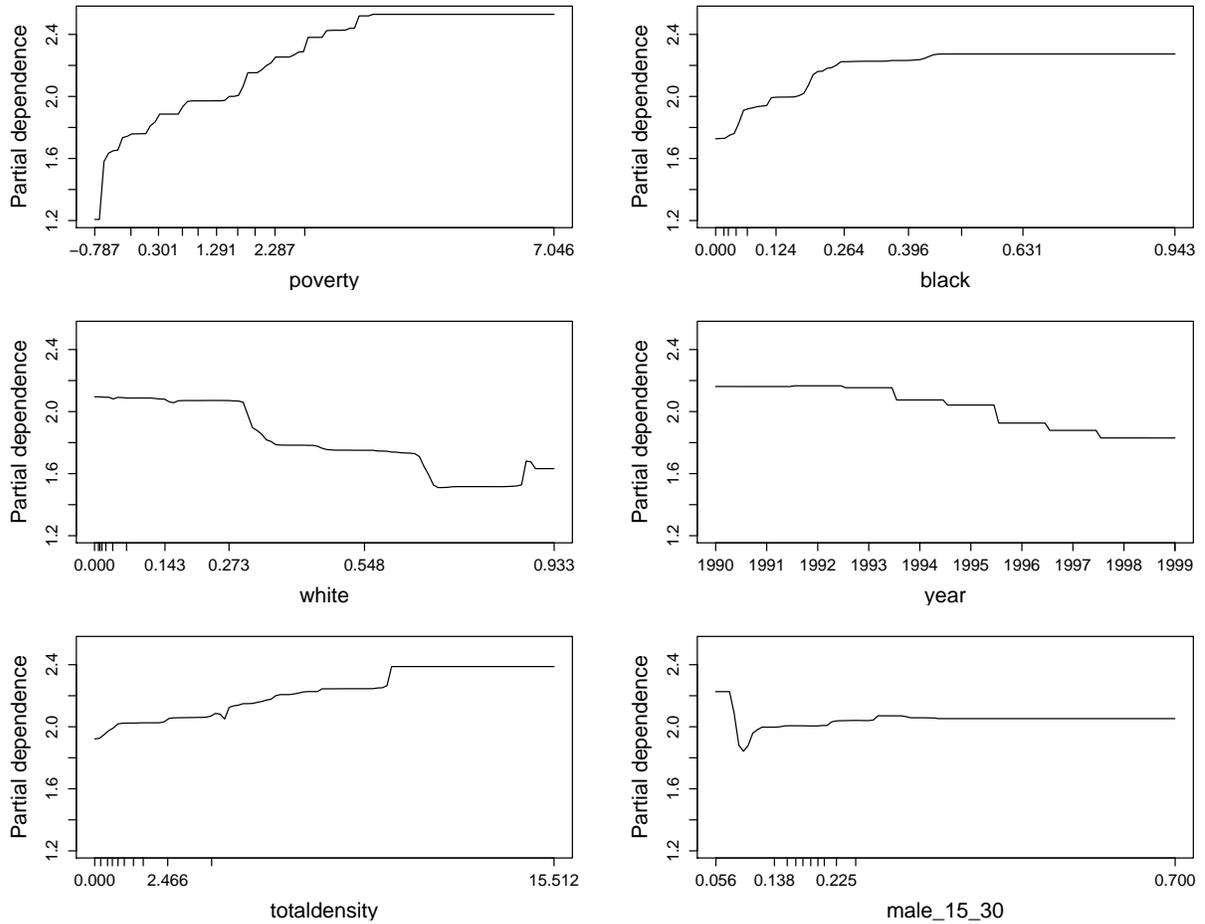}
 \caption{\it{Partial dependence plots in MART-fitted model A.}}\label{fig-par}
\end{figure}

After establishing the relative importance of the explanatory variables, the nature of the dependence of the fitted model on any subset of explanatory variables is of interest.  The partial dependence function can help us to graphically examine the dependence of a fitted model on low cardinality subsets of the variables.  Figure~\ref{fig-par} based on model A shows the partial dependence plots for the first six most important variables.  We see that higher alcohol outlet density is associated with higher assault rates.  Also the assault rates were decreasing over the years from 1990 to 1999 with a steeper decreasing after 1993.  The assault rates decreasing with the percentages of young males in the tracts and then increasing a little and then stabilized.  Furthermore, census tracts with relatively more blacks as well as poorer tracts, tend to have higher risks.

\begin{figure}[htb]
 \centering
 \includegraphics[angle=270,width=0.8\textwidth]{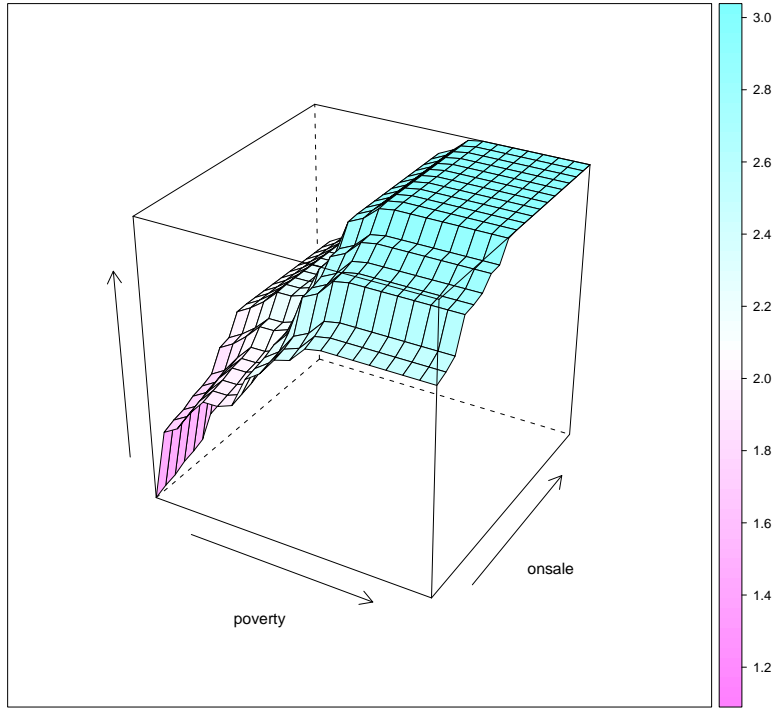}
 \caption{\it{Two dimensional partial dependence plot of {\it poverty} and {\it onsale} in MART-fitted model B.}}\label{fig-par2}
\end{figure}

Figure~\ref{fig-par2} shows the two-dimensional partial dependence plot of {\it poverty} and {\it onsale} in MART-fitted model B. We see that both {\it poverty} and {\it onsale} act positively on the response with no obvious interaction pattern. Friedman and Popescu (2005) developed techniques which allow us to test the total interaction strength for each input variable. The procedure is essentially a variant of permutation test. For details of the test procedure, we refer the readers to the Section 8 in Friedman and Popescu (2005). We applied the procedure to the fitted MART Model A \& B. No significant interaction effects is observed for all the predictor variables. 

\subsection{The spatial correlations} 

In this section, all analysis are based on model A.  To check the spatial heterogeneity explained by our models, Table~\ref{moran} shows the Morans' I, the index of spatial correlation, of the original log assault rate, the residuals after fitting with covariates and the residuals after both spatial smoothing and covariate fitting over the ten years.  Note that after MART fitting with covariates, the remaining residuals in the years 1991 and 1996 have no significant spatial correlation, thus the two years of data are not used in the second stage to fit the spatial errors.  We find that most spatial correlations are explained by the hierarchical additive model.  There are still spatial correlations in the years 1990, 1992 and 1999.  But the spatial correlation becomes negative.  It seems that in these years, the spatial associations are overfitted.  Note that these years are the years when the p-values of spatial tests are relatively large after covariates fitting, meaning less spatial correlation remained in the residuals.  Also note that the year 1992 is when the civil unrest happened.  

\vspace{0.5cm}
\renewcommand{\baselinestretch}{1.0}
\begin{table}[htb]
\centering
\begin{center}
\doublespacing
  \begin{tabular}[]{l|rrrrrr} \hline\hline
    Year  & Origin & P-value & Res1 & P-value & Res2 & P-value \\ 
\hline
1999 & 0.46 &     0 & 0.13 &  0.00 & -0.10 & 0.01 \\
1998 & 0.52 &     0 & 0.29 &  0.00 & -0.05 & 0.21 \\
1997 & 0.50 &     0 & 0.25 &  0.00 & -0.03 & 0.46 \\
1996 & 0.29 &     0 & 0.05 &  0.06 & 0.05 & 0.06 \\
1995 & 0.49 &     0 & 0.23 &  0.00 & -0.03 & 0.49 \\
1994 & 0.49 &     0 & 0.24 &  0.00 & -0.05 & 0.17 \\
1993 & 0.52 &     0 & 0.21 &  0.00 & -0.02 & 0.68 \\
1992 & 0.30 &     0 & 0.14 &  0.00 & -0.16 & 0.00 \\
1991 & 0.18 &     0 & 0.02 &  0.41 & 0.02 & 0.41 \\
1990 & 0.44 &     0 & 0.16 &  0.00 & -0.11 & 0.00 \\ \hline \hline
  \end{tabular}
  \caption{\it{Moran's I and P-values of Spatial Correlation Testing: the ``Origin'' column is the Moran's I of the original log assault rate; ``Res1'' is the moran's I of the remained residuals after the covariates fitting; and ``Res2'' is the moran's I of the remained residuals after the two-stage fitting.  The P-values of the spatial correlation tests are shown in the right column to the corresponding Moran's I.}}
  \label{moran}
\end{center}
\end{table}

\begin{figure}[htb]
 \centering
 \includegraphics[angle=0,width=0.9\textwidth]{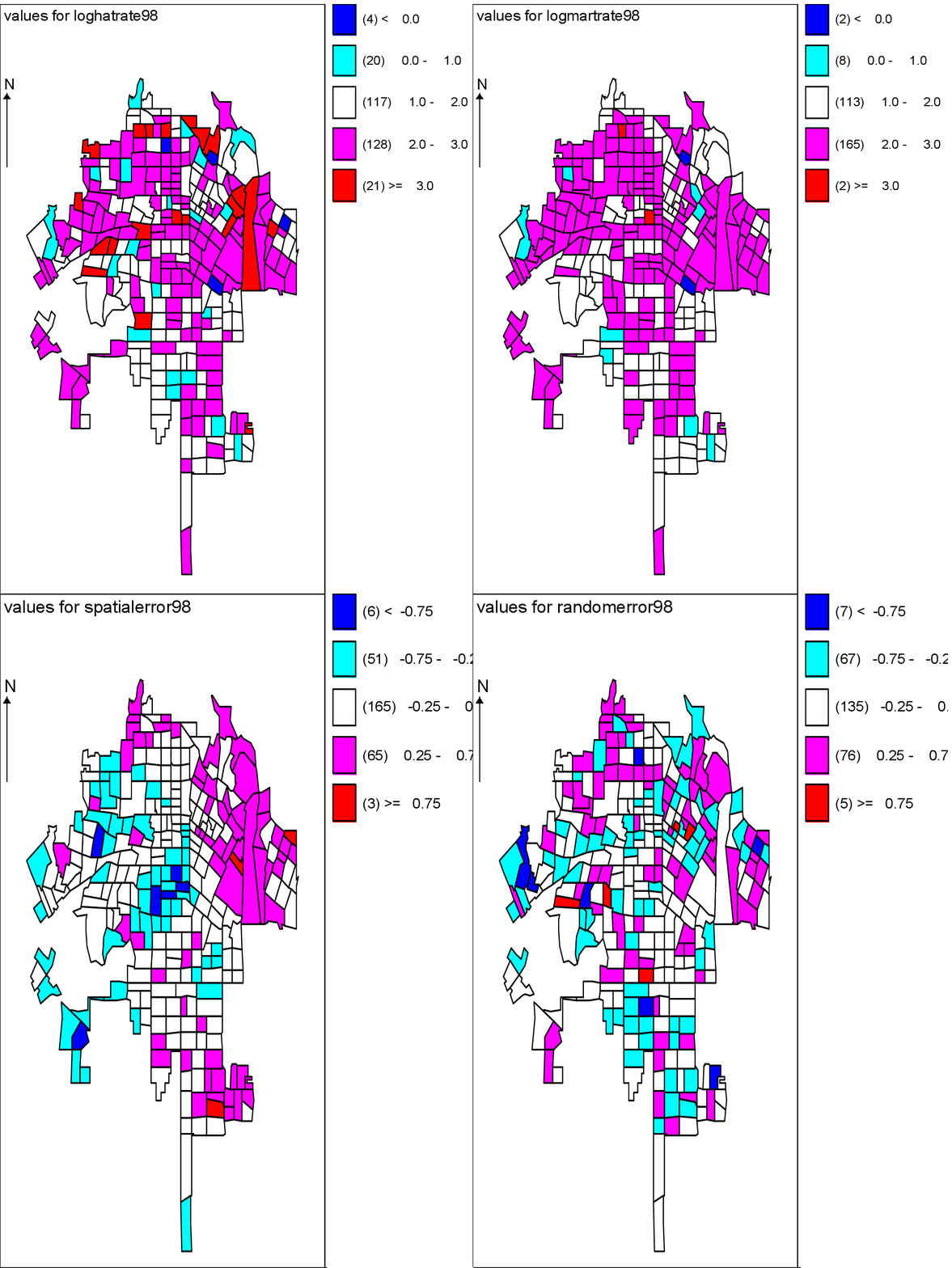}
 \caption{\it{Maps of log assault rates on fitted data. The left upper map is the two-stage fitted assault rate in 1998, and the right upper panel maps the MART fitted log assault rates.  The left lower panel maps the fitted spatial error and the right lower panel shows the random residuals after taking out the spatial errors. }}\label{map98_single}
\end{figure}

Figure~\ref{map98_single} maps the distribution of the fitted assault rates in 1998 in the civil unrest area.  The upper panels separately map the fitted log assault rates using the hierarchical additive models and those fitted through MART only.   The left upper panel reveals a concentration of assaults in the middle to east areas, which comprise downtown LA and its immediate neighborhoods.   The lower panels map residuals, where the left panel is the fitted spatial error ($\phi_{1998,C_i}$) and the right panel is the random residuals after all the model fitting, i.e., the raw log assault rate minus the log fitted assault rate at each census tract in 1998.  We see no obvious spatial correlations remained in the residuals after the two-stage analysis.  The remaining spatial autocorrelation in the residuals after covariates fitting of MART suggests the presence of unmeasured spatially varying covariates.  One obvious candidate here would be the distance of each tract from the origin of the civil unrest (the intersection of Florence and Normandy).  Additional possible factors are more social in nature, and are related to the fact the southern region of the civil unrest area tends to be the most disadvantaged in the city.  While measures of SES and ethnicity are already included in the model, other factors associated with concentrated disadvantage (i.e., family structure, ethnic isolation, low social capital) were not and may contribute to the observed patterns.  Maps from other years can also be drew and analyzed to figure out possible missing variables.

\subsection{Comparison of methods}

Yu et al. (2007) analyze the same data set using hierarchical linear regression models, in which some covariates and interactions are explored and transformed into linear models.  The spatial errors are also modeled with CAR strategy in their paper.  Several models are compared, with the best model chosen as the final model and used to explain the effects of interesting.  The model is chosen in terms of the DICs (Spiegelhalter et al., 2002) of the model.  Here we compare their Bayesian hierarchical linear model with our hierarchical additive model in terms of the spatial correlations explained by models and the model explanation.

Still, we use the Moran's I to do the spatial correlation test.  The Moran's I and the test p-value of the remained residuals after model fitting with the hierarchical linear model is shown in Table~\ref{yu-moran}, which should be compared with the last two columns of Table~\ref{moran}.  We see that there are still lots of spatial correlations remained in the residuals after the hierarchical linear regression.  The hierarchical additive model does a better job in explaining the spatial correlations in this data set.

\vspace{0.5cm}
\renewcommand{\baselinestretch}{1.0}
\begin{table}[htb]
\centering
\begin{center}
\doublespacing
  \begin{tabular}[]{l|rrrrrrrrrr} \hline\hline
    Year  & 1999 & 1998 & 1997 & 1996 & 1995 & 1994 & 1993 & 1992 & 1991 & 1990 \\ 
\hline
Moran's I & 0.14 & 0.24 & 0.18 & 0.04 & 0.13 & 0.20 & 0.17 & 0.11 & 0.02 & 0.16 \\
p-value   & 0.00 & 0.00 & 0.00 & 0.11 & 0.00 & 0.00 & 0.00 & 0.00 & 0.39 & 0.00\\ \hline \hline
  \end{tabular}
  \caption{\it{Moran's I and P-values of Spatial Correlation Testing For the Residuals form the Hierarchical Linear Model.}}
  \label{yu-moran}
\end{center}
\end{table}

We also notice that the pD, explained as the effective number of model parameters (Spiegelhalter et al., 2002), from the hierarchical linear model is 2356, while that from the hierarchical additive model is only 873.  Both pDs account for the local shrinkage of the spatial random effects only.  This means that most variances in the assault rates are explained through the spatial errors in the linear model, while the hierarchical additive model more efficiently uses the covariates to explain the assault rates.  

Yu et. al (2007) find that compared with the census tracts that had no off-premise alcohol license surrender in the 1992 civil unrest, the census tracts that had off-premise alcohol license surrender experienced a steeper drop in assaultive violence rate one year after the civil unrest, with the effect lasting roughly five years.  The hierarchical additive model fails to recognize this effect since MART can only find out relatively more important covariates and interactions, while the hierarchical linear model is built by using human knowledge efficiently: covariates and interactions of great interest are forced to enter the model and tested for significance, thus suggesting a ``wiser'' use of our hierarchical additive model: we could transform the variables and create interactions according to our previous knowledge and use that to fit a final model. 

\section{Conclusions and Future Works}\label{conclusion}

In this paper, we propose a hierarchical additive model strategy - using nonparametric method to build the relationship among variables and utilizing a CAR model to smooth the spatial heterogeneity.  We use this strategy to explore the relations between alcohol availability and assault rates.  We have demonstrated that the total alcohol outlet density is positively related to the assault rate, also that the on-premise alcohol outlets are even more important than the off-premise alcohol outlets in predicting assault rate in Los Angeles.  We capitalized on the natural experiment of 1992 Los Angeles civil unrest but did not find important consequences on assault rates from the alcohol license surrender.  Maps were provided to show the distribution of fitted assault rates, as well as residual maps to suggest possible missing covariates.  Our method has been compared with the hierarchical linear model and showed superior performance in exploring important variables in explaining the change of assault rates.

As mentioned above, many other variables could have been included in our model and some lagged effect from coefficients could also be used to smooth the remaining spatial correlations.  In addition, it might be of greater interest to model other types of crimes with assaults simultaneously.  Our future research would to analyze different alcohol related assault violences such as assault, homicide, rape and robbery together.  A possible solution is to use MART to explore the relationships between different violence and the covariates separately and then use the multivariate intrinsic Gaussian CAR hyper-distribution on the variance terms to explore the remained residuals together.  This could be easily realized through the ``mv.car'' function in Winbugs.  A final area of interest is to study the association between alcohol availability and mortality rates in the study region using the hierarchical additive model.

\end{document}